\documentclass{iopart}
\usepackage{iopams,graphicx}
\begin{document}
\title[Scattering of light and SPPs by indentations]{Theory on the scattering of light and surface plasmon polaritons
by arrays of holes and dimples in a metal film.}
\author{F de Le\'on-P\'erez$^1$, G Brucoli$^1$, F J Garc\'ia-Vidal$^2$ and L Mart\'in-Moreno$^1$}

\address{$^1$Instituto de Ciencia de Materiales de Arag\'{o}n and Departamento de F\'{i}sica de la Materia Condensada,
CSIC-Universidad de Zaragoza, E-50009, Zaragoza, Spain}
\address{$^2$Departamento de F\'{\i}sica Te\'{o}rica de la Materia Condensada, Universidad Aut\'{o}noma de Madrid, E-28049 Madrid, Spain}
\ead{lmm@unizar.es}
\begin{abstract} The scattering of light and surface plasmon polaritons (SPPs) by finite arrays of
either holes or dimples in a metal film is treated theoretically. A
modal expansion formalism, capable of handling real metals with up to
thousands of indentations, is presented. Computations based on this method
demonstrate that a single hole scatters a significant
fraction of incoming light into SPPs. It is also observed that holes
and dimples scatter
SPPs into light with similar efficiencies, provided the depth of the dimple is
larger than its radius.
Finally, it is shown that in arrays the normalized-to-area emittances in the
out-of-plane and SPP channels present different dependences 
with the number of holes.

\end{abstract}
\vspace{2pc}

\submitto{\NJP}

\section{Introduction}

\begin{figure}
 \centering
\includegraphics[width=\columnwidth]{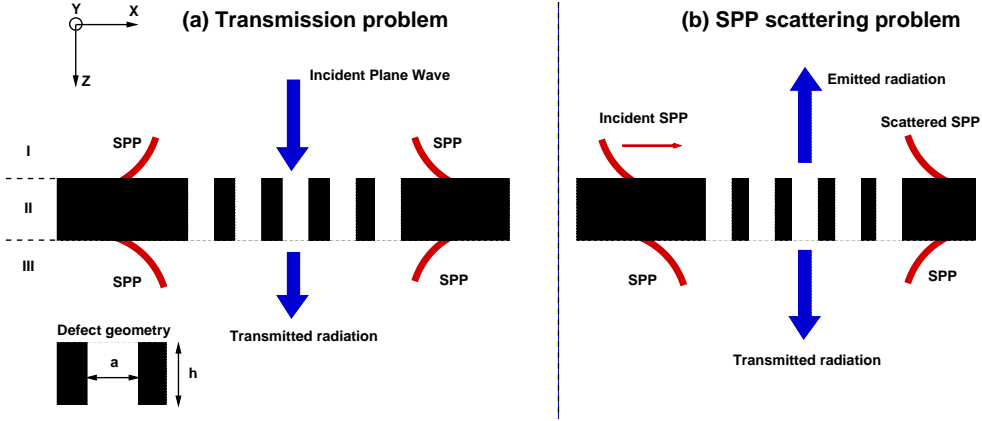}
\caption{(Color). Schematic lateral view of the
system under study, its different scattering mechanisms, and the frame of reference used. In panel (a) the system is
back-illuminated by a plane wave. The relevant scattering channels under illumination by a SPP are represented in panel (b).}
 \label{fig:scheme}
\end{figure}

Surface plasmon polaritons (SPPs) are electromagnetic modes bound to a metal-dielectric interface \cite{Raether}. Their unique optical properties have,
among other applications, the potential for designing highly integrated
photonic circuits with length scales much smaller than those
currently achieved.  Such circuits should convert light into SPPs,
propagate them to logic elements where they are to be processed and
ultimately converted back into light, as discussed in
\cite{BarnesN03}.

The theoretical analysis of these processes is hampered by the
difficulty of computing the optical properties of nanostructured
metals. Standard techniques capable of providing virtually exact
results to Maxwell's equations can nowadays only treat either very small
systems or systems with a high degree of symmetry. In this way, optical transmission through a single hole has been
studied with the Multiple Multipole Method \cite{WannemacherOC01}, Finite Difference
Time Domain (FDTD) simulations \cite{ChangOE05}, the Green Dyadic approach \cite{Colas2005,Bortchagovsky2006,SepulvedaOE08}, and special
methods devised for treating circular holes \cite{PopovAO05} or 1D systems (slits) \cite{LalannePRL05}.
Optical transmission through {\it periodic infinite} arrays of holes
have been studied with FDTD \cite{BaidaPRB03,RodrigoPRB08}. 

Nevertheless,
approximate methods are required when computing either the
optical transmission through finite arrays of holes, or the
scattering coefficients of SPPs impinging on finite structures. 
Regarding the scattering properties of SPPs, 
several approaches have been developed in order to study
collections of 1D scatterers (i.e. with translational
symmetry in one direction)\cite{SanchezGilAPL98,SanchezGilAPL05,FLTNP07,NikitinPRB08}. 
In the 2D case (holes,
protrusions, dimples...) the available theoretical formalisms are
based on the coupling of dipoles, an approximation that is only valid when the dimensions of the scatterers are much smaller than the
wavelength (specially when, as usual, the
polarizability of the scatterer is represented by its quasi-static
value)\cite{EvlyukhinSS05,deAbajoPRL05,LaluetOE07}.

Our first main goal is to present a method for treating
the electromagnetic properties of up to thousands of {\it
indentations} in a real metal. We apply this
formalism for computing (i) how much energy goes into the
different SPP channels when an array of holes is back illuminated
(see \fref{fig:scheme}(a)), and (ii) the scattering coefficients of
SPPs by finite arrays of 2D indentations (see \fref{fig:scheme}(b)).

The indentations can be either holes
or dimples (i.e. holes closed at one end), with arbitrary
size and shape, placed at arbitrary positions. Our approach is
based on a modal expansion of the fields in different spatial
regions. It is a non-trivial extension from the formalism previously
developed for dealing with either
1D indentations in a real metal \cite{FLTPRB05} or 2D indentations in a perfect electrical conductor
(PEC), where EM fields can not penetrate \cite{BravoPRL04}. The model we shall present has already been applied to computing the plasmon coupling
efficiency of a single defect (both with circular and rectangular
shape) \cite{baudrionOE08}. Furthermore, it was used for calculating the optical transmittance through
different systems: a
single rectangular hole \cite{FJPRB06}, a periodic array of
rectangular holes \cite{MaryPRB07}, and a finite array of circular
holes \cite{przybillaOE08}. Excellent agreement 
was obtained in all these cases with both experimental results
\cite{baudrionOE08,przybillaOE08} and 
FDTD simulations \cite{FJPRB06,MaryPRB07,przybillaOE08} (when
available). However, no detailed account of its derivation has been presented before.

Our second main goal in this
paper is to analyze the results on launching and decoupling of SPP by metal gratings
reported in \cite{DevauxAPL03}. These experiments motivate our
choice of geometrical parameters.

The paper is organized as follows. \Sref{sec:me} is devoted to presenting the theoretical framework used in the manuscript. In order to facilitate the
reading, in \Sref{sec:me} we give an overview of the method
and define the quantities studied throughout the
paper. The derivation of the formalism as well as several useful
expressions are given in the appendixes. 
In \sref{sec:singledefect} we present the optical response of
a single circular hole (illuminated by either a plane wave or a
SPP), while in \sref{sec:metalgrating} we look at the scattering
coefficients for arrays of circular holes.

\section{Modal expansion formalism}
\label{sec:me}

We consider a set of indentations (either holes or dimples) of
arbitrary shape, and arbitrarily placed in a planar metal film
(infinite in the x-y plane and having finite thickness $h$). The
system can be divided in the three regions shown in
\fref{fig:scheme}. Region I and region III are dielectric
semi-spaces characterized by the real dielectric constant
$\epsilon_1$ and $\epsilon_3$, respectively. Region II represents
the corrugated metal film with a wavelength-dependent dielectric
function $\epsilon_M$. Holes or dimples could also be
filled with a dielectric $\epsilon_2$. We assume that the system is
illuminated by EM wavefields coming from region I.

We expand the EM fields on the eigenmodes of each region, and match them at the boundaries.
The finite dielectric constant of the metal is taken into account by using
Surface Impedance Boundary Conditions (SIBCs) \cite{jackson}.
Roughly speaking, the results obtained with the SIBCs
can be understood as a second order Taylor expansion in  $z_s=1/\sqrt{\epsilon_M}$.
The first order term ($z_s=0$) is the result obtained 
within the PEC approximation, and was reported in \cite{BravoPRL04}. The use of SIBCs does not simply represent a quantitative improvement over results obtained with the PEC approximation: it also allows for the computation of scattering of
SPPs (which do not exist in flat PEC interfaces but already appear when SIBCs
are employed). Notice however that SIBC can not describe tunneling of EM fields directly across the metal. Therefore our method is only applicable to metal thickness larger than 2-3 skin depths, when direct tunneling is negligible. A cautionary remark: when dimples are considered, this restriction in thickness should be applied from the bottom of the dimple rather than from the metal surface.

\begin{figure}
 \centering
\includegraphics[width=12cm,height=6cm,angle=0,bb=50 150 749 450]{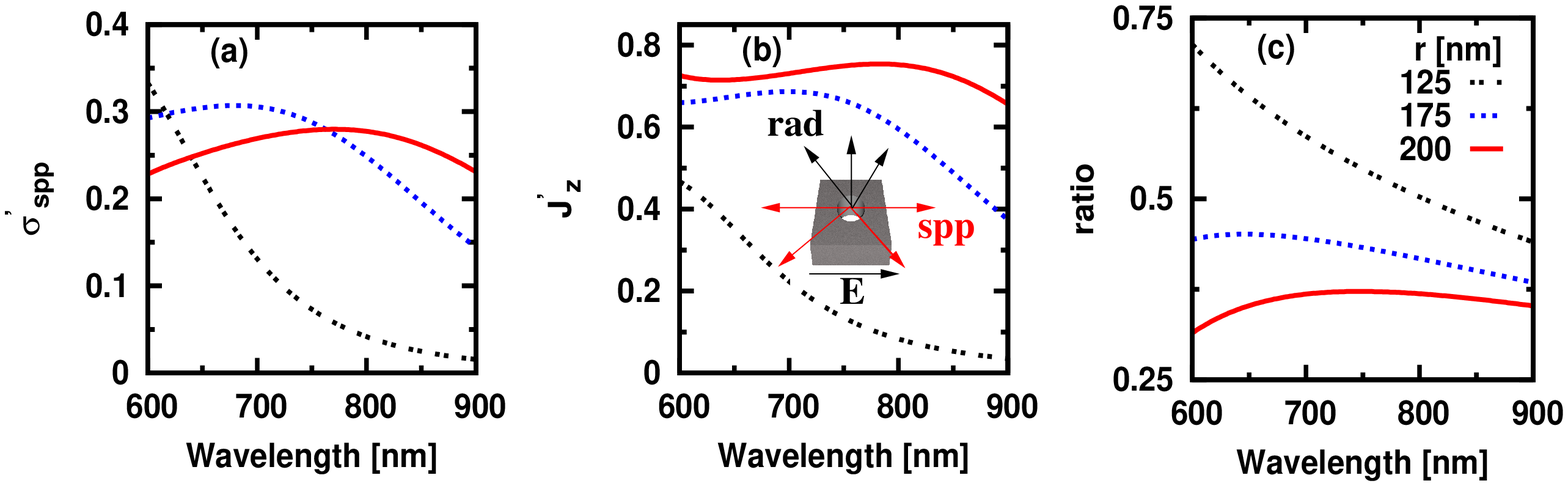}
\caption{(Color). Optical properties of circular hole in a gold film
illuminated by a normal-incident plane wave, for different hole
radii $r$. The metal thickness is fixed to $h$=150. The system
under study is schematically represented in the inset to panel (b),
illustrating the two relevant scattering channels: radiation and
launching of SPP. Panel (a) shows the spectral dependence of the
normalized-SPP-emittance $\sigma'_{\scriptscriptstyle spp}$, while Panel
(b) renders the normalized-transmittance $J'_z$. The spectral
dependence on the ratio $\sigma'_{\scriptscriptstyle spp}/J'_z$ is shown in panel (c).}
 \label{fig:circlevsr}
\end{figure}

After some algebraic manipulations we end up with the following coupled system
of equations for $E_{\alpha}$ and $E'_{\alpha}$, which are essentially
the modal amplitudes of the electric field
at the input and output interfaces of the indentations, respectively
($\alpha$ is an index labeling each waveguide mode in each indentation considered in the calculation).

\begin{eqnarray}
\label{eq:SE0}
\begin{array}{l}
 \left( G_{\alpha \alpha}-\Sigma_{\alpha}\right)  E_{\alpha}+\sum_{\beta \neq \alpha} G_{\alpha \beta} E_{\beta} -G^V_{\alpha} E'_{\alpha} = I_{\alpha} \\ \\
\left( G'_{\gamma \gamma}-\Sigma_{\gamma}\right)  E'_{\gamma}+\sum_{\nu \neq \gamma} G'_{\gamma \nu} E'_{\nu} -G^V_{\gamma} E_{\gamma}= 0.
\end{array}
\end{eqnarray}

Details of the derivation, as well as the expressions for the different
quantities involved can be found in \ref{app:OSE}. Let us just
mention that the ingredients required for the calculation are
the spatial dependence of waveguide modes in each indentation, and
their propagation constants. These modes are analytically known
for some hole geometries; otherwise they can be numerically computed
solving a 2D problem. In this section we just give the
physical meaning of the different quantities in \eref{eq:SE0}. $I_{\alpha}$
represents the direct illumination over waveguide
mode $\alpha$. It is proportional to the overlapping of the incident
electric field with the mode in the indentation. The rest of the terms
take into account the (self-consistent) wandering of fields between the
indentations. $\Sigma_{\alpha}$ is related to the bouncing back and
forth of a given waveguide mode inside an indentation, due to the
discontinuity that the waveguide mode faces at the end of the indentation. The main
difference between holes and dimples is the presence of
$G^V_{\alpha}$, which only appears for holes, and is related to the
EM field on one side of the hole due to the presence of an EM field
on the other side. The coupling between different waveguide modes is
given by the "propagator" $G_{\alpha \beta}$. This takes into
account that the EM field emitted by each point within object $\beta$
can be ``collected'' by the object $\alpha$. The propagator $G'_{\gamma
\nu}$  differs from $G_{\alpha \beta}$ in the constituent parameters
only, i.e.  $G'_{\gamma \nu}$ is a function of $\epsilon_3$ while
$G_{\alpha \beta}$ depends on $\epsilon_1$.

Notice that this formalism is reminiscent of the Green Dyadic
approach \cite{MartinPRL95}. In that method the EM fields must be computed in the
volume {\it inside} the indentations. In contrast, our method provides
the EM fields everywhere in terms
of the fields {\it at the openings} of the indentations. This very
compact representation allows for the treatment of the optical
properties of systems involving a large number of indentations, 
at a relatively low
computational cost \cite{przybillaOE08}.

\begin{figure}
\centering
\includegraphics[width=12cm,height=6cm,angle=0,bb=50 100 749 450]{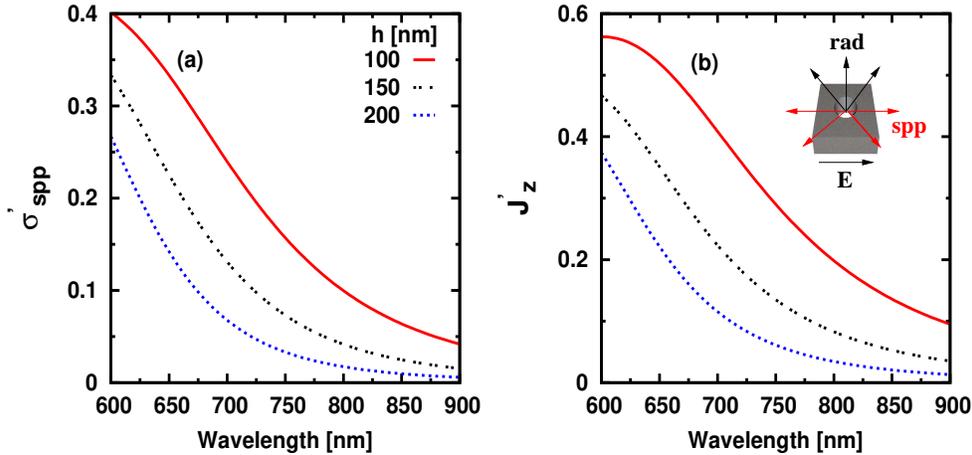}
\caption{(Color). Optical properties of a circular hole in a Au film
illuminated by a normal-incident plane wave, for different metal
thicknesses $h$. The hole radius is $r$=125 nm. The system under study
is schematically represented in the inset to panel (b). Panel (a)
shows the spectral dependence of the normalized-SPP-emittance
$\sigma'_{\scriptscriptstyle spp}$, while Panel (b) renders the
normalized-transmittance $J'_z$.} \label{fig:circlevsh}
\end{figure}

The price we pay is that the modal expansion formalism is approximate and
relies on the SIBCs, while the Green Dyadic method is virtually exact. 
The implementation of modal expansion calculations usually faces 
another source of inaccuracy. 
The required overlaps between plane waves and waveguide modes 
are only known for some waveguide geometries {\em defined in a PEC}. 
Nevertheless with some minor corrections, waveguide 
modes in a PEC can still yield good agreement with
experimental results (and exact calculations, when available). 
For instance, using the propagation constant of the 
waveguide in a real metal greatly improves the prediction
of the spectral position of resonances. 
Another improvement is to enlarge  the hole
side, so as to simulate the real penetration of the field at
the lateral wall defining the waveguide. Doing so, our method has provided
good agreement with the FDTD simulation for the optical transmission through a single rectangular hole \cite{FJPRB06}. In that case, the hole was enlarged as proposed in
\cite{GordonOE05}. 

In this paper we will only consider circular indentations,
for which the propagation constant of waveguide modes 
in a real metal can be easily computed \cite{straton}.
The expressions for the waveguide modes are borrowed from those of a
circular waveguide in the PEC \cite{straton,RobertsJOSA87}. The radius of the hole is phenomenologically enlarged by one
skin depth. Such enlargement provides the best agreement
with FDTD simulations for an infinite periodic array of
holes \cite{przybillaOE08}.

As said before, once the coefficients $E_{\alpha}$ and
$E'_{\alpha}$ are known, it is possible to obtain the fields everywhere. In
particular, the energy power scattered into
different channels can be computed by integration of the relevant
time-averaged components of the Poynting vector. We
define $W_z$ as the total radiated power crossing a fictitious plane
placed at constant z in region I. The definition of $W^{\scriptscriptstyle II}_z$ and $W'_z$ is the same as $W_z$, except the appropriate fictitious plane is placed in region II and III, respectively. Notice that we are maintaining the notation used in
the system of equations, where primed quantities refer to region III.
The values of these quantities in terms of the set
$E_{\alpha}$ and $E'_{\alpha}$ can be found in \ref{sec:TW}. In a
real metal, EM energy power can also leave the system as a SPP wave. We
define  $W_{\scriptscriptstyle spp}$ and $W'_{\scriptscriptstyle
spp}$ as the energy power in the SPP channel that crosses an ideal
cylinder with axis parallel to the $z$ direction, placed in either
region I or III, respectively. Details and expressions for these
quantities can be found in \ref{sec:SW}.

We shall present results in terms of {\it normalized}
emittances $J$, defined as the total power $W$ divided by both the
incident power impinging onto the area covered by indentations in region I,
$W_{inc}$, and the total number of holes in the system $N$.
In other words, for each scattering channel we define
a corresponding emittance as $J=W/(W_{inc} N)$, independently
of whether the incident field is a SPP or a radiation field.
However, when illumination is via a SPP, 
$W_{inc}$ is defined as the total power crossing the infinite 
imaginary strip 
perpendicular to the metal surface, whose base coincides with the
maximum geometrical cross section of the collection of indentations.

For a lossy system the total scattered power in the SPP channel $W_{\scriptscriptstyle spp}$ \eref{eq:Wspp} is a function of the observation point, $R$, at the metal surface. As plasmon's decay with in-plane
distance is well known, it is possible to define the total scattered power in the plasmon
channel {\it as if the plasmon were excited at $R=0$}. We denote this quantity by $\Omega_{\scriptscriptstyle spp}$ \eref{eq:Ospp}, which satisfies
$W_{\scriptscriptstyle spp}=\Omega_{\scriptscriptstyle spp} \exp (-2 \vert \mbox{Im} [k^{\scriptscriptstyle \rm spp}_{\parallel} ] \vert R)$ \eref{eq:Wspp}.
Nevertheless, the comparison of efficiencies into the SPP channel between systems
of different sizes is not trivial in lossy systems.  We need to treat separately the case of back-illuminated and SPP illuminated arrays. 

\begin{figure}
 \centering
\includegraphics[width=12cm,height=6cm,angle=0,bb=50 100 749 450]{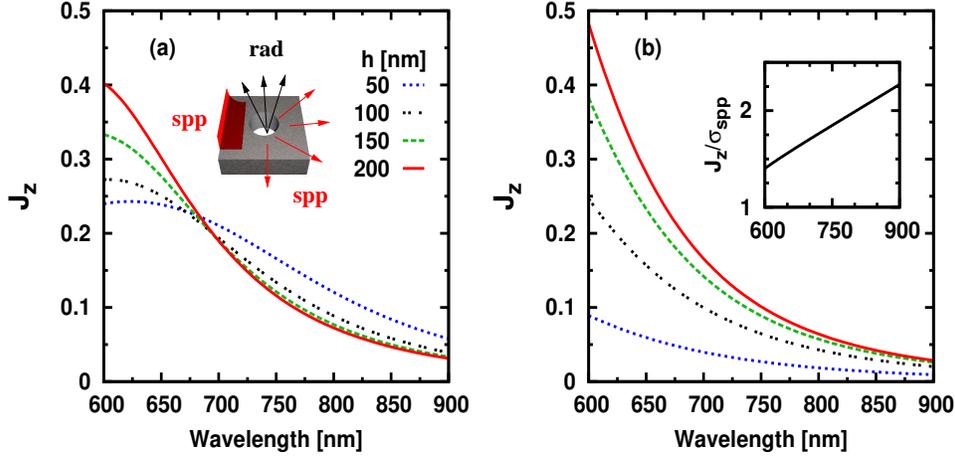}
\caption{(Color). Spectral dependence of the normalized
out-of-plane emittance, $J_z$, for a single circular indentation in
gold illuminated by a SPP: panel (a) circular hole, panel (b) circular dimple.  In both
cases the radius of the indentation is $r$=125 nm and different
indentation depths $h$ have been considered. Inset to panel (a)
shows a scheme of the system. Inset to panel (b) renders the
spectral dependence of the ratio $J_z/\sigma_{\scriptscriptstyle spp}$
(which is the same function for both holes and dimples and for
different $h$).} \label{fig:singledefectiluspp}
\end{figure}

When comparing efficiencies 
into the SPP channel for back-illuminated arrays of different sizes at fixed $R$, larger systems may have emitters of
SPPs closer to that reference distance $R$, leading to exponentially larger values of $\Omega_{\scriptscriptstyle spp}$ than those of smaller arrays. In order to avoid this geometric artifact in the definition of $\Omega_{\scriptscriptstyle spp}$, we
analyze the SPP field at a fixed distance $R_e$ from the edge of the array. 
If we measure distances from the center of the array: $R=R_e + L_{sys}$, where we define  $L_{sys}$ as the distance from the center of the array to its edge. Then the normalized SPP emittance is 
$J_{\scriptscriptstyle spp}= \sigma_{\scriptscriptstyle spp} 
\exp(-2 \vert \mbox{Im} [k^{\scriptscriptstyle \rm spp}_{\parallel} ] 
\vert R_e)$, 
with $\sigma_{\scriptscriptstyle spp}=\Omega_{\scriptscriptstyle spp} / (W_{inc} N) \exp(-2 \vert \mbox{Im} [k^{\scriptscriptstyle \rm spp}_{\parallel} ] \vert L_{\scriptscriptstyle sys})$. The "cross-section" $\sigma_{\scriptscriptstyle spp}$ is the quantity we focus on throughout the paper, as it
provides the normalized intensity into the SPP channel {\it on a circle passing through the edge of the array}.

When the array is illuminated by a SPP, however, using the same convention may 
lead to spurious size dependences in the "cross-section" $\sigma_{\scriptscriptstyle spp}$. It is not convenient to place the origin on the center of the array since the SPP wave may not even reach
the center for large arrays and strong SPP scattering. Hence, we place the origin at the center of the first column of
indentations encountered by the incident SPP and analyze the response of different 
arrays at the same distance $R$ from this origin, irrespectively of the array size. Therefore, in this case, we have $J_{\scriptscriptstyle spp}= \sigma_{\scriptscriptstyle spp} \exp(-2 \vert \mbox{Im} [k^{\scriptscriptstyle \rm spp}_{\parallel} ] \vert R)$, with $\sigma_{\scriptscriptstyle spp}=\Omega_{\scriptscriptstyle spp} / ({W_{inc} N})$. 

Let us stress that
the previous conventions are chosen in order to obtain a cross-section into the SPP channel which allows for the comparison between arrays of different sizes; however, these conventions do not affect neither 
the total flux $W_{\scriptscriptstyle spp}$ nor 
the flux normalized to both area and number of holes $J_{\scriptscriptstyle spp}$. 

Let us close this section by reminding that current conservation imposes relations that are useful for checking
the correctness of both the derived expressions and the computer
codes. For instance, if absorption is neglected, the identity
$J^{II}_z=J'_z+J'_{\scriptscriptstyle spp}$ should be fulfilled, i.e
the fraction of the incident energy traversing the hole, $J^{II}_z$,
must be either radiated into freely propagating waves, $J'_z$, or
scattered into SPP at the output side of the metal film,
$J'_{\scriptscriptstyle spp}$. On the input side we have a similar
relation, which now depends on whether the incident field is a plane
wave or a SPP. Although all results presented in this paper include absorption, we have checked that for a lossless metal our code provides current conservation up to a relative error of 0.1\% of the incident flux.

\section{Single defect}
\label{sec:singledefect}

\begin{figure}
 \centering
\includegraphics[width=12cm,height=6cm,angle=0,bb=50 100 749 450]{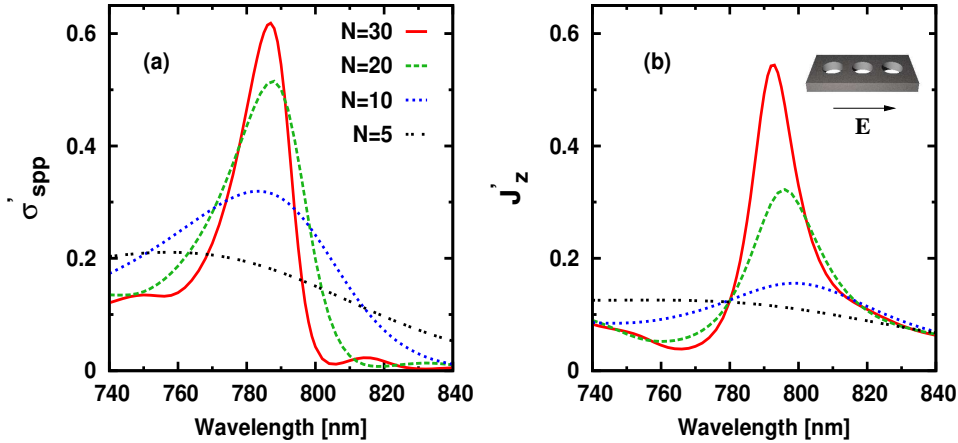}
\caption{(Color). Spectral dependence of both
$\sigma'_{\scriptscriptstyle spp}$ (a) and $J'z$ (b) for a "horizontal"
linear array of circular holes with radius $r$=125 nm, $h$=150 nm,
and period P=760 nm, for different array sizes. The system under
study, schematically represented in the inset, is back-illuminated by a plane wave.}
 \label{fig:lineararray}
\end{figure}

In this section we shall study the optical properties of a single
circular hole in a real metal, when illuminated by either a plane
wave or a SPP. We shall analyze both the out-of-plane radiation and
how much energy goes into the plasmon channel.

Let us first consider back-illumination of the hole by a plane wave,
impinging at normal incidence. The optical transmittance through a
single hole perforated in a real metal has been largely studied both
experimentally
\cite{baudrionOE08,przybillaOE08,ThioOL03,DegironOC04,YinAPL04} and
theoretically
\cite{WannemacherOC01,ChangOE05,PopovAO05,FJPRB06,ShinPRB05}. In
general, the transmittance is characterized by Fabry-P\'erot peaks, one of them
being very broad and appearing close to the cutoff wavelength of the
fundamental waveguide mode \cite{FJPRB06,ShinPRB05}. The previously
cited theoretical works have looked at the spatial dependence of scattered fields, finding that a hole launches
SPPs. However, to the best of our knowledge, no computation has yet been performed on the
efficiency of the SPP launching by a back-illuminated hole (Ref.
\cite{baudrionOE08} considered the coupling into the SPP channel
{\it at the interface of incidence}). We have performed such calculation,
analysing the scattered EM arising from the SPP pole in the Green dyadic (see \ref{sec:SW}). This analysis is
possible even in the presence of absorption, although in lossy metals the actual EM fields at the metal surface decay with the
distance from the hole due to the dissipation of SPP energy into heat.

\Fref{fig:circlevsr} renders the (normalized-to-area) fraction
of incident energy that is scattered by a circular hole into
either the SPP ($\sigma'_{\scriptscriptstyle spp}$, panel (a))
or radiation ($J'_z$, panel (b)) channels. Both quantities are computed in the region of
transmission. The metal considered is gold, whose dielectric constant is fitted to Palik's data \cite{Palik}. The film thickness is $h$=150 nm, and
different hole radii are studied.
For the sake of simplicity we assume a
freestanding metal film since the peak investigated in the
experiment concerns the metal-air interface. As \fref{fig:circlevsr}
shows, both $\sigma'_{\scriptscriptstyle spp}$ and $J'_z$ present
non-monotonous spectral dependencies, due to a broad resonance appearing close to the cutoff
wavelength of the fundamental waveguide mode, $\lambda_c$. For Gold,
we obtain $\lambda_c$(r=125 nm) = 589 nm, $\lambda_c$(r=175 nm) =
732 nm, and $\lambda_c$(r=200 nm) = 811 nm. Notice that spectra are plotted from $\lambda=$ 600 nm, because far from $\lambda <600$ nm results are not reliable because the SIBC approximation breaks down(i.e $\sqrt{\epsilon_m(\lambda)}\ll 1$ is not longer fulfilled in that regime). However, in the considered spectral window, as much as 30\% of
the energy impinging into the holes can be converted into SPPs.

\Fref{fig:circlevsr} (c) shows that the
ratio $\sigma'_{\scriptscriptstyle spp}/J'_z$ is
a smooth radius-dependent function of wavelength. For the parameters
considered, this ratio is of the order of 0.3-0.8, being larger for
smaller holes. In the long-wavelength limit ($\lambda \gg r$) this ratio can be
worked out analytically; we find that
$\sigma'_{\scriptscriptstyle spp}/J'_z \sim \vert z_s \vert \sim
\lambda^{-1}$ for a metal represented by
a Drude dielectric constant. This decrease of the coupling to SPPs with wavelength
originates from the weakening of the SPP confinement to the surface,
which translates into a weaker coupling to indentations.

In our formalism the ratio $\sigma'_{\scriptscriptstyle
spp}/J'_z$ is independent of metal thickness. However, both
$\sigma'_{\scriptscriptstyle spp}$ and $J'_z$ depend on $h$, as shown in
\fref{fig:circlevsh} for a single hole with $r=125$.
$\sigma'_{\scriptscriptstyle spp}$ and $J'_z$
decay with both $h$ and $\lambda$ in the spectral window considered, 
where all fields inside the hole are evanescent.  

Let us consider now the decoupling of SPPs into radiation,
after they have been scattered by either
a hole or a dimple (see scheme in \fref{fig:scheme}(b)). 
To the best of our knowledge, this problem is virtually unexplored. We are only aware of a study on SPP scattering by a shallow
dimple with a Gaussian profile \cite{MaradudinPRL97}.

\Fref{fig:singledefectiluspp} shows the spectral dependence of
the energy power radiated into the region of incidence, $J_z$, when a SPP
impinges into either a circular hole (a) or a circular dimple (b). In
both cases the radius is $r=$125 nm. Several values of $h$ are considered
($h$ is the metal thicknesses in the
case of a hole, and the depth of the indentation in the case of a
dimple). The inset of \fref{fig:singledefectiluspp}(b) renders the
ratio $J_z/\sigma_{\scriptscriptstyle spp}$ (notice that this is
the inverse of the quantity shown in \fref{fig:circlevsr}(c)). This ratio
depends on the radius of the indentation, but does not depend
neither on $h$ nor on whether the indentation is a hole or a dimple.
For the considered case this ratio is always larger than unity,
implying that a single hole scatters better into radiation than back
into SPPs. \Fref{fig:singledefectiluspp} also shows that a hole and a
dimple couples SPPs into radiation with similar intensity, provided
the indentation is deep enough. Dimples may provide higher values of $J_z$
than holes, due to the lack of
radiation into region III.  However, shallow holes radiate more than
dimples. This is trivial when $h\rightarrow 0$ since the dimple then disappears. We find that, in the subwavelength
regime, $h\approx r$ is a rough rule
of thumb for the depth at which the scattering of SPPs into radiation is 
the same for dimples and holes of the same size. For smaller $h$, conversion of
SPPs into radiation is substantially smaller
for a dimple than for a hole. Notice though
that this "transition" depth depends on the evanescent decay of the
fields inside the indentation, and therefore on wavelength. For instance,
a hole illuminated by a SPP at $\lambda=$800 nm
radiates 35\% more than a dimple, when both have $h=$150 nm.

\section{Array of circular holes}
\label{sec:metalgrating} 
We shall now investigate scattering by arrays of
circular holes of both a normal incident
plane wave and a SPP.

\begin{figure}
 \centering
\includegraphics[width=12cm,height=6cm,angle=0,bb=50 100 749 450]{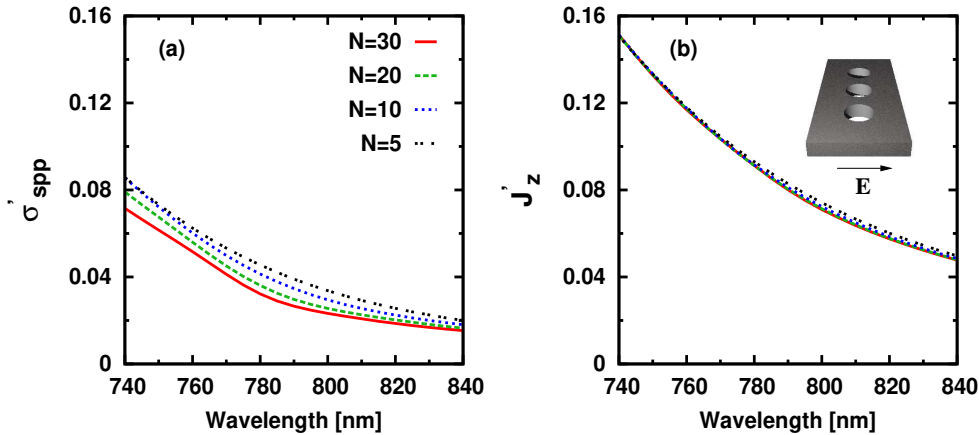}
\caption{(Color). Spectral dependence of both
$\sigma'_{\scriptscriptstyle spp}$ (a) and $J'z$ (b) for a "vertical"
linear array of circular holes with radius $r$=125 nm, $h$=150 nm,
and period P=760 nm, for different array sizes.
The system under
study, schematically represented in the inset, is back-illuminated by a plane wave.}
 \label{fig:verarray}
\end{figure}

Let us first focus on the transmittance and the
launching of SPP via back-illumination of the array. In the perfect conductor
approximation, the normalized-to-area optical transmission through linear
arrays of holes shows resonances of the same order as those in square
arrays \cite{BravoPRL04}. This occurs if the array is lined up {\it parallel} to the incident electric
field. In contrast, the normalized-to-area optical transmission
through an array of holes {\it perpendicularly oriented} to the
incident field is very similar to that of a single hole. Here we
consider these configurations and analyze whether the same phenomenology
holds for the launching of SPPs.

\begin{figure}
 \centering
\includegraphics[width=12cm,height=6cm,angle=0,bb=50 100 749 450]{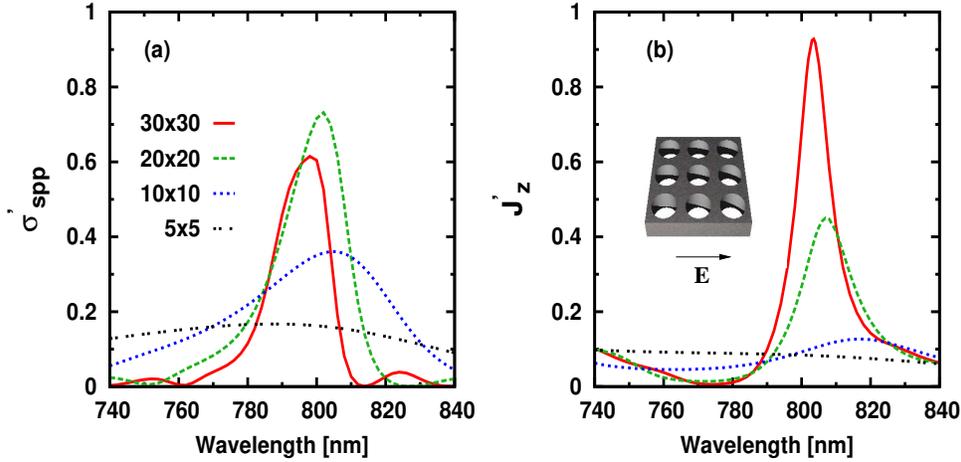}
\caption{(Color). Spectral dependence of both
$\sigma'_{\scriptscriptstyle spp}$ (a) and $J'z$ (b) for a square array
of circular holes with radius $r$=125 nm, $h$=150 nm, and period
P=760 nm, for different array sizes. The system under
study is schematically represented in the inset.}
 \label{fig:sqrarray}
\end{figure}

Take a set of holes with radius $r$=125 nm, separated by a distance $P$= 760 nm in a
metal layer of thickness $h$=150 nm (again, the geometrical parameters of \cite{DevauxAPL03}).
\Fref{fig:lineararray}, \fref{fig:verarray}, and \fref{fig:sqrarray} render
the results for a linear array lined up parallel to the
incident electric field ("horizontal" array),
a perpendicularly aligned one ("vertical" array), and a square array, respectively. In all three figures, panel (a) shows the spectral dependence of the
normalized-to-area SPP emittance, $\sigma'_{\scriptscriptstyle spp}$,
while panel (b) renders the normalized-to-area transmittance $J'_z$, for a different number of holes.

Clearly,
holes in the "vertical" configuration (\fref{fig:verarray}) are virtually
independent. Conversely scattering of light by "horizontal" arrays (\fref{fig:lineararray}) shows resonances due to interaction between holes. These resonances also appear
in the square arrays (\fref{fig:sqrarray})\footnote{Notice that, in order to
better resolve the details close to the peaks, the spectral window
shown here is smaller than the one used when presenting results for
the single hole}. Notably, the wavelengths of SPP emittance peaks appear slightly redshifted
with respect to the periodicity. Actually, the position of the
SPP emittance peak is also slightly redshifted with respect to the
wavelength of SPPs  {\it of the flat metal interface} with in-plane
wavevector $k^{\scriptscriptstyle spp}=2 \pi / P$ 
(this wavelength is $\lambda
= 777$ nm for gold and period $P=760$ nm). This redshift reflects the fact that the EM field spends some time
inside the indentations, slightly modifying the resonant conditions obtained by considering  geometrical arguments only.

\begin{figure}
 \centering
\includegraphics[width=\columnwidth,height=6cm,bb=50 100 410 250]{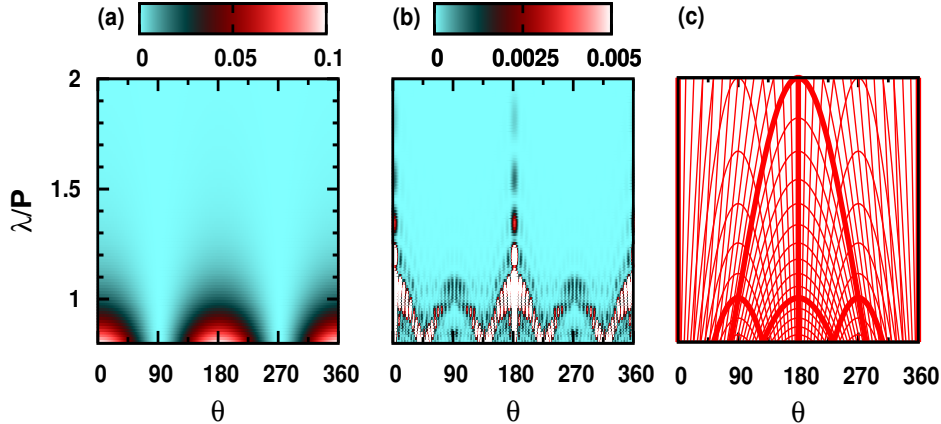}
\caption{(Color). Angular and wavelength dependence of the emittance
of SPPs, $\sigma'_{\scriptscriptstyle spp}(\lambda,\theta)$, by a
back-illuminated set of holes. The incident field is a plane wave
with electric field pointing along one of the cartesian axis of the
array, direction from which $\theta$ is measured (in degree). Panel (a):
single circular hole, panel (b): $10
\times 10$ square array of circular holes with $P$=760 nm.
In both cases $r=$150 nm and $h=$150 nm. Panel (c) renders the
curves of maxima (heavy line) and minima (thin lines) intensity for
a diffraction grating of point emitters \cite{born}.}
 \label{fig:jpangle}
\end{figure}

At resonance, the
normalized-to-area transmittance $J'_z$ is larger in the square array than 
in the "horizontal" one. Additionally, the
normalized-to-area emittance $\sigma'_{\scriptscriptstyle spp}$ has
already reached saturation for arrays with 20x20 holes, while it has not saturated
yet for "horizontal" arrays with $N=30$ holes. This different
behavior with number of holes can be understood by analyzing the
angular dependence of the power scattered into SPP,
$\sigma'_{\scriptscriptstyle spp}(\lambda,\theta)$ ($\theta=0 $ coincides
with the direction of the incident electric field). Panel (a) of
\fref{fig:jpangle} renders a contour plot $\sigma'_{\scriptscriptstyle
spp}(\lambda,\theta)$ for an isolated circular hole in Au ($r$=125
nm, $h$=150 nm). As the figure illustrates, a hole launches SPPs mainly in
the directions parallel and antiparallel to the incident electric
field. This is because a SPP is partly a longitudinal
wave, with its in-plane wavevector parallel to the in-plane electric
field. The SPP launched by a "horizontal" array of holes at
angles that are not close to either $\theta=0$ or $\theta=180$
degree are not further scattered and thus contribute to
$\sigma'_{\scriptscriptstyle spp}$. In a square array, however, those
"off-axis" SPPs which are launched by holes close to the center of the array
can be scattered by other holes. This scattering can be
either into SPPs or into radiation. Such process increases $J'_z$ and
decreases $\sigma'_{\scriptscriptstyle spp}$, and
explains the different size dependencies of "horizontal" and square
arrays.

The angular dependence of the SPP emittance for a square array of
10x10 holes, is shown in \fref{fig:jpangle}(b). The hole radius and
metal thickness are those of the single hole case of \fref{fig:jpangle}(a), while the period is $P$= 760 nm. The angular dependence of the SPP
emittance by the array can be understood as the one due to the
single hole, modulated by the interference pattern characteristic of
a diffraction grating \cite{born}. We recall that, for light
impinging normally on a finite "vertical" diffraction grating
composed of $N$ point emitters, intensity maxima
appear at angles $\theta$ given by the relation $\lambda/d=2
\sin^2(\theta/2) n/N$ where both $n$ and $n/N$ are integers. Angles
of intensity minima satisfy the same equation with $n$ integer and
$n/N$ non-integer. Similarly, finite "horizontal"
gratings satisfy $\lambda/d=2\vert  \sin \theta \vert n/2N$, with the same conditions as the previous case governing the presence of maxima and minima in terms of $n/N$.
The maxima of both horizontal and vertical gratings are represented
by heavy lines in \fref{fig:jpangle}(c), while the minima are
represented by thin lines. The emittance of SPP by a hole array (\fref{fig:jpangle}(b))
combines both the pattern of maxima and minima
observed in \fref{fig:jpangle}(c) and the single hole emittance (\fref{fig:jpangle}(a)).

\begin{figure}
 \centering
\includegraphics[width=12cm,height=5cm,bb=50 120 410 232]{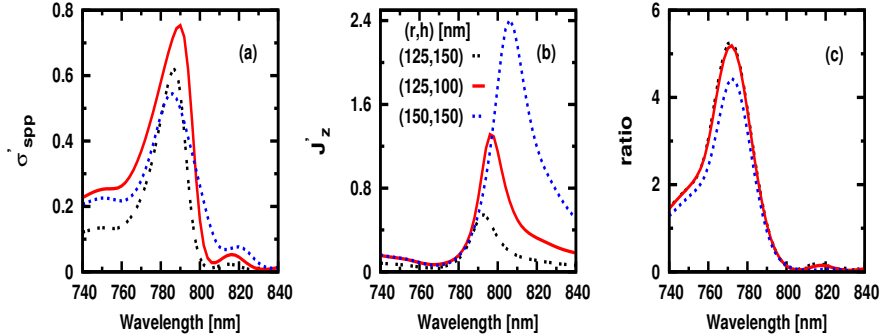}
\caption{(Color). Spectral dependence of
$\sigma'_{\scriptscriptstyle spp}$ (a), $J'_z$ (b), and the ratio $\sigma'_{\scriptscriptstyle spp}/J'_z$  for a "horizontal"
linear array of 30 circular holes with period P=760 nm and varying radius $r$ and metal thickness $h$. The system under study is
back-illuminated by a plane wave. }
 \label{fig:lineararrayvshandr}
\end{figure}

The influence of hole geometry on the SPP and radiation efficiencies in arrays 
is illustrated in \fref{fig:lineararrayvshandr}, which renders the emittance spectra for a "horizontal"
linear array of 30 circular holes with period P=760 nm as function of both its radius $r$ and metal thickness $h$. The hole array with $r$=125 nm and $h$=150 nm, already studied in \fref{fig:lineararray}, is compared with both an array of smaller $h$ and the same $r$ and an array of larger $r$ and the same $h$.  Reducing the film thickness to $h$=100 we increase the total transmission through the system increasing the amount of energy scattered in both channels, $\sigma'_{\scriptscriptstyle spp}$ in \fref{fig:lineararrayvshandr}(a) and $J'_z$ in \fref{fig:lineararrayvshandr}(b), keeping the the same radio as for $h$=150, see \fref{fig:lineararrayvshandr}(c). On the other hand, when the circle radius is increased to $r$=150 nm the transmittance of the system is favored against  $\sigma'_{\scriptscriptstyle spp}$, leading to a lower relative ratio. Notice, however, that these properties may present different dependences with number of holes for different hole sizes. This point deserve further investigation which exceeds the scope of the present paper. 

\begin{figure}
 \centering
\includegraphics[width=\columnwidth,height=6cm,bb=50 100 410 250]{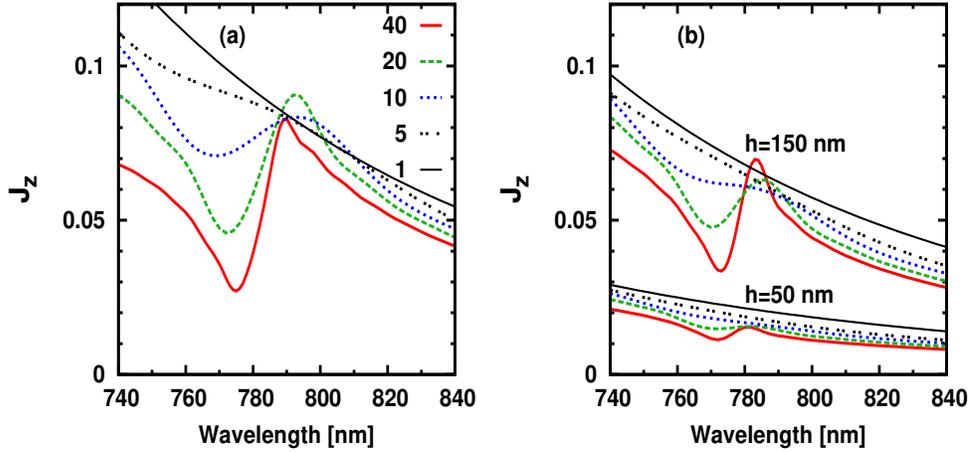}
\caption{(Color). Normalized emittance, $J_z$, as function of the
wavelength, for a "horizontal" linear array of indentations illuminated by a SPP.
(a) hole array with $h=$150 nm. (b) dimple array for both $h=$150
nm and $h$=50 nm. In all cases, $r$=125 nm, $P=$760 nm and different array sizes
have been considered. The single hole spectra is included to facilitate the comparison.}
 \label{fig:arrayhvsd}
\end{figure}

Let us now focus on the scattering of an incident SPP by a
"horizontal" linear array of indentations. As seen, the
properties of a square array of indentations are very similar to
those of "horizontal" arrays. It has been experimentally found \cite{DevauxAPL03}
that square arrays of holes (with
radius $r=$125 nm drilled in a gold film 150nm thick) decouple SPPs
into radiation, while dimples with the same radius and 50 nm depth
do not radiate any measurable EM signal. In both cases the period of
the array was $P=$760 nm. \Fref{fig:arrayhvsd} presents 
the spectral dependence
of the normalized out-of-plane
radiation, $J_z$, for "horizontal" arrays (holes: panel (a),
dimples: panel (b)). The same geometrical parameters as in the
experiments were used. Additionally, in \fref{fig:arrayhvsd}(b) we
present results for dimples with the same depth as that of the holes considered in the
experiment. For the geometrical parameters
and spectral window considered, this figure shows that, $J_z$ develops a resonance, which appears
slightly redshifted from the period. As
\fref{fig:arrayhvsd} shows, arrays formed with either holes or
dimples with $h$=150 nm scatter SPP into radiation with similar
efficiencies. However, in line with the experimental results, the
computed intensity of this process for arrays of dimples with $h$=50
nm is around five times smaller, at the wavelength considered in the
experiment ($\lambda=800$ nm).
In any case, notice that for these geometrical parameters, for which
scattering by a single indentation is weak, the normalized out-of-plane radiation is of the same order for arrays and for single indentations 
(results for the single hole of \fref{fig:singledefectiluspp} are included in \fref{fig:arrayhvsd} 
in order to facilitate the comparison).

\begin{figure}
 \centering
\includegraphics[width=\columnwidth,height=6cm,bb=50 100 410 250]{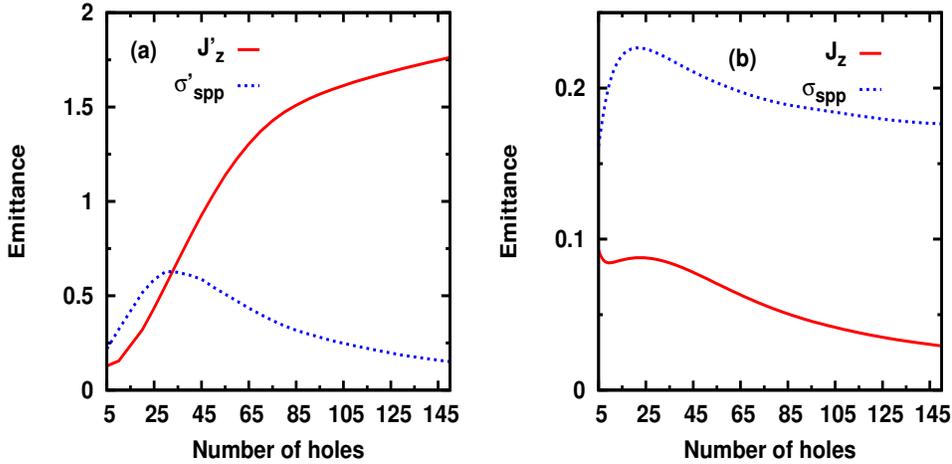}
\caption{(Color). Normalized out-of-plane and SPP emittance
at resonance for a "horizontal" linear array of circular holes (with
radius $r$=125 nm, $h$=150 nm, and period P=760 nm), as function of
array size. (a) Back-illumination by a plane wave. (b) Illumination
by a SPP.}
 \label{fig:horpeakvsn}
\end{figure}

As we have seen, "horizontal" arrays of holes produce resonant
behavior in both the launching of SPPs and the radiated energy. This
occurs independently of whether the array is illuminated with a plane wave or by a SPP. However, all these efficiencies
present very different dependencies with the size of the array. In
order to illustrate this point we consider "horizontal" arrays of
holes, with $r$=125 nm and $P=$760 nm, drilled in a gold film of
thickness $h=$150 nm. \Fref{fig:horpeakvsn}(a) renders the
evolution of the maximum value of both
$\sigma'_{\scriptscriptstyle spp}$ and $J'_z$ with the number of holes (at the resonance around
$\lambda\approx 790$nm), when the array is back-illuminated by a
plane wave.  This figure completes  the analysis of the peaks in 
\fref{fig:lineararray} considering arrays of larger sizes. For the geometrical parameters analyzed,
$\sigma'_{\scriptscriptstyle spp}$ reaches a maximum for arrays of about 35
holes. In contrast, the saturation of $J'_z$ has not yet been reached even
for arrays of 150 holes. In \Fref{fig:horpeakvsn}(b) we plot the size dependence of both the maximum of $J_z$ and the maximum of
$\sigma_{\scriptscriptstyle spp}$ when the "horizontal" array is illuminated by a SPP. We recall that these quantities are
normalized to the total number of holes. In this case, for large
arrays, both $J_z$ and $\sigma_{\scriptscriptstyle spp}$ 
decrease with the size of the array.
This is because part of the incident SPP
current is radiated as it progresses along the array, therefore
weakening the illumination of the holes at the further end of the
array. The values of $J_z$
are even smaller in "horizontal" arrays than in single holes. Consequently, 
"vertical" arrays (which, as we have seen behave as a collection of
independent holes) are a better choice for out-coupling of SPPs
than "horizontal" arrays.

As result of the hole-hole interaction, we find that in a hole array at resonance 
$Jz>\sigma_{\scriptscriptstyle spp}$, in contrast with the
result for the single hole. This behavior can be understood in terms of simple physical arguments. In-plane emittance for a single hole occurs essentially along the forward-backward direction, see \fref{fig:jpangle}(a). Along this direction, at resonance, the EM fields of the individual holes are in phase, interfering constructively. Therefore, the total EM field along these directions of favored SPP emission increase with number of holes. 
On the other hand, the out-of-plane single hole emission (not shown here) is practically isotropic. Therefore, the interference of the individual holes is either constructive or destructive, depending on the optical path to a given observation point. As a result of having both constructive and destructive interference, the total contribution to this scattering channel is smaller than for the in-plane radiation. 

It is important to emphasize that the emittances reported in \fref{fig:horpeakvsn} are measured at resonance. This explains why the
figure considers $N\geqslant 5$: for smaller $N$ the peak is practically not yet developed, see \fref{fig:lineararray} and  \fref{fig:arrayhvsd}. 

\section{Summary}
We have presented a theoretical formalism for dealing with optical properties of sets of indentations in
an optically thick metal film. The indentations may
have arbitrary shape and can be arbitrarily placed at any
of the two metal interfaces.
The formalism is based on a modal expansion of the fields in the different
regions of space. Fields are matched by
Surface Impedance Boundary Conditions. These boundary conditions take into
account the dielectric constant of a real metal and can therefore be used
to approximately describe surface plasmon polaritons. Hence, our
formalism can be used for analyzing optical problems like
transmission through
hole arrays, scattering of SPP by sets of indentations and launching of
SPP by back-illuminated arrays, among others. Our systematic study of the optical properties of both circular isolated
indentations and
arrays of indentantions in gold has found that:
\begin{itemize}
\item Close to the cutoff condition, a back-illuminated hole scatters
a significant fraction of light into SPPs in the region of transmission ($\approx 30\%$).
\item A hole and a dimple of the same size
scatter SPPs into radiation with comparable
efficency if, roughly speaking, the depth of the indentation
is not smaller that its radius.
Otherwise, holes couple SPP into radiation more efficiently.
\item Linear arrays oriented along the direction of the impinging
electric field exhibit resonances
similar to those appearing in square arrays of holes.  This holds in the coupling into both  radiation and SPPs. Furthermore, the appearance of these resonances is independent of whether the illumination is via a plane wave or a SPP. 
\item  Linear arrays oriented perpendicularly to
the direction of the impinging electric field
essentially behave as collection of independent holes.
\item Illuminating by a SPP or a plane wave results in a substantially different behavior of the emittance as function of the array size. 
\item The SPP channel saturates  for smaller number of holes than the radiation channel, when a hole array is illuminated by a plane wave.
\end{itemize}

\ack{We gratefully acknowledge financial support from the European Network of Excellence Plasmo-Nano-Devices (FP6-2002-IST-1-507879) and the Spanish MICINN under contract  MAT2005-06608-CO2, Grant No. AP2005-55-185, and Juan de la Cierva program.}

\appendix
\setcounter{section}{0}
\section{Derivation of the modal expansion 
formalism with Surface Impedance Boundary Conditions.}
\label{app:OSE} We solve Maxwell's equations 
(written in CGS units) for the electric, $ \bi{E}$, and magnetic, $\bi{H}$, fields in a system comprising a planar metal film
corrugated with a set of indentations. 
The film is in the x-y plane and has a finite
thickness $h$.
 The indentations can be either holes or dimples 
(i.e. holes that do not pierce through the film),
with arbitrary shape and placed at arbitrary positions at both input
and output interfaces.

We assume the system lies in a rectangular supercell, with lattice
parameters $L_x$ and $L_y$ along the $x$ and $y$ axes, respectively.
This supercell may be real (if we are considering a bona fide
periodic system) or artificial, if the number of defects is finite.
In the latter case, the limit $L_x,L_y\rightarrow \infty$ must be
taken. The space within the supercell is divided into the three
regions shown in \fref{fig:scheme}. Region I and region III are
dielectric semi-spaces characterized by the real dielectric constant
$\epsilon_1$ and $\epsilon_3$, respectively. Region II (which extends from $z=0$ to $z=h$) represents
the corrugated metal film with a wavelength-dependent dielectric
function $\epsilon_M$. Holes or dimples could be also filled with a
dielectric $\epsilon_2$. We assume the EM energy incident on the
metal layer is coming from region I (see \fref{fig:scheme}).

In order to take advantage of known solutions of
Maxwell's equations the EM fields of the constituent media are
represented in a convenient basis. We use Dirac's notation, such that $\left\langle \bi{r_{\parallel}}\vert\bi{E}\right\rangle =
\bi{E}(\bi{r_{\parallel}})=(E_x(\bi{r_{\parallel}}),E_y(\bi{r_{\parallel}}))^t$,
where t stands for transposition and $\bi{r_{\parallel}}=(x,y)$.

In region I, see \fref{fig:scheme}, the fields are
expanded into an infinite set of plane waves with parallel
wavevector $\bi{k}_{\parallel} =(k_x,k_y)$ and polarization
$\sigma$. We consider that the incident radiation
(labeled by the superscript $0$)  is either a
plane wave or a SPP. Arbitrary illumination can be easily considered by
decomposing first the incident wavepacket into plane waves, and subsequently
integrating the optical
response of the array to each plane wave.
The fields can be written in terms of the reflection
amplitudes $\rho_{\bi{k}_{\parallel} \sigma}$ as
\begin{eqnarray}
\label{eq:EMFI}
\fl \begin{array}{l}
 \vert \bi{E} > = \rme^{\rmi k^0_z z} \vert \bi{k}^0_{\parallel} \sigma^0>+\sum_{\bi{k}_{\parallel} \sigma} \rho_{\bi{k}_{\parallel} \sigma} \rme^{-\rmi k_z z} \vert \bi{k}_{\parallel} \sigma>, \\ \\
\vert -\mathbf{u_z} \times \bi{H} > = Y_{\bi{k}^0_{\parallel} \sigma^0} \rme^{\rmi k^0_z z} \vert  \bi{k}^0_{\parallel} \sigma^0>-\sum_{\bi{k}_{\parallel} \sigma} Y_{\bi{k}_{\parallel} \sigma} \rho_{\bi{k}_{\parallel} \sigma} \rme^{-\rmi k_z z} \vert \bi{k}_{\parallel} \sigma >.
\end{array}
\end{eqnarray}
where $\mathbf{u_z}$ is unitary vector directed along $z$-direction and the summation runs over wavevectors of the form $\bi{k}_{\parallel} = \bi{k}^0_{\parallel} +
\bi{K}_\bi{R}$, $\bi{K}_\bi{R}$ being a vector of the supercell reciprocal lattice.
The real space representation of the plane
waves is given by
\begin{equation}
\label{eq:PW}
 \left\langle r_{\parallel} \vert \bi{k}_{\parallel} \sigma \right\rangle=\frac{\exp(\rmi \bi{k}_{\parallel} \cdot r_{\parallel})}{k_{\parallel} \sqrt{L_x L_y }} \cases{ (k_x,k_y)^t, & p-polarization \\
(-k_y,k_x)^t, & s-polarization }
\end{equation}
where we assume a $\bi{k}_{\parallel} \sigma$-independent
normalization $\left\langle \bi{k}_{\parallel}
\sigma \vert \bi{k}_{\parallel} \sigma \right\rangle=1$. The
electric and magnetic fields in \eref{eq:EMFI}  are related through
the admittance $Y_{\bi{k}_{\parallel} p}=k_\omega \epsilon_1/k_z$ and
$Y_{\bi{k}_{\parallel} s}=k_z/k_\omega$ where $k_\omega=2\pi / \lambda$
($\lambda$ is the wavelength of the incident radiation), and
$k^2_{\parallel}+k^2_z=\epsilon_1 k^2_\omega$.

In the transmission region (region III in \fref{fig:scheme}, where $z\geq h$) the EM fields are also
expanded in plane waves, and expressed in terms of the
transmission amplitudes $t_{\bi{k}_{\parallel} \sigma}$
\begin{eqnarray}
\label{eq:EMFIII}
\begin{array}{l}
 \vert \bi{E}' > = \sum_{\bi{k}_{\parallel} \sigma} t_{\bi{k}_{\parallel} \sigma} \rme^{\rmi k'_z (z-h)} \vert \bi{k}_{\parallel} \sigma>, \\ \\
\vert -\mathbf{u_z} \times \bi{H}' > = \sum_{\bi{k}_{\parallel} \sigma} Y'_{\bi{k}_{\parallel} \sigma} t_{\bi{k}_{\parallel} \sigma} \rme^{\rmi k'_z (z-h)} \vert \bi{k}_{\parallel} \sigma>.
\end{array}
\end{eqnarray}
Note that all quantities in region III are primed in order to
distinguish them from those in region I.

Waveguide modes are
the most natural choice for expanding the EM field inside the
indentations. The main ingredients needed in the present formalism are both
the propagation constants and the EM fields of waveguide modes in all indentations.
In general, these quantities are attained by solving
an EM problem with reduced (cylindrical) symmetry. For holes with a given symmetry (circular, rectangular...) these modes
are known analytically, if the surrounding metal is considered a perfect electrical conductor.
In such favorable situation, the penetration of the EM field into the real metal can be
taken into account by phenomenologically enlarging the hole.

 Writing  the fields as function of the expansion
coefficients $C_{\alpha}$ and $D_{\alpha}$, we have
\begin{eqnarray}
\label{eq:EMFII}
\begin{array}{l}
 \vert \bi{E}^{\scriptscriptstyle II} > = \sum_{\alpha} \left(C_{\alpha} \rme^{\rmi k_{\alpha z} z}+D_{\alpha} \rme^{-\rmi k_{\alpha z} z} \right) \vert {\alpha} >, \\ \\
\vert -\mathbf{u}_z \times \bi{H}^{\scriptscriptstyle II}> = \sum_{\alpha} Y_{\alpha} \left(C_{\alpha} \rme^{\rmi k_{\alpha z} z}-D_{\alpha} \rme^{-\rmi k_{\alpha z} z} \right) \vert {\alpha} >,
\end{array}
\end{eqnarray}
where $\alpha$ labels a certain waveguide mode inside a certain
aperture. In what follows we shall refer to a given element of the
set $\lbrace \alpha \rbrace$ as an object. The propagation constant
along the $z$ direction of eigenmode $\alpha$ is given by
$k_{\alpha z}$, while $Y_{\alpha}=k_{\alpha z}/k_\omega$ and
$Y_{\alpha}=\epsilon_2 k_\omega/k_{\alpha z}$ are the admittances for TE
and TM modes, respectively.

In order to take into account the dielectric properties of a real metal, we impose
surface impedance boundary conditions (SIBCs) \cite{jackson}. SIBCs establish the following relation between
the tangential components of the EM fields at the horizontal interfaces: $\bi{E}=z_s \bi{H}
\times \bi{u}_n$, where $\bi{u}_n$ is the vector normal to the
surface and pointing to the interior of the metal and $z_s=1/\sqrt{\epsilon_M}$.

The previous continuity equation must be satisfied at each interface.
Projecting over the different plane waves, at both $z=0$ and $z=h$, we get

\begin{eqnarray}
\label{eq:rhok}
\begin{array}{l}
\rho_{\bi{k}_{\parallel} \sigma}=-\frac{f^-_{\bi{k}_{\parallel}
\sigma}}{f^+_{\bi{k}_{\parallel} \sigma}} \delta_{\bi{k}_{\parallel},\bi{k}^0_{\parallel}} \delta_{\sigma, \sigma^0}\, + \,  \frac{1}{f^+_{\bi{k}_{\parallel} \sigma}}\sum_{\alpha} \left\langle  \bi{k}_{\parallel} \sigma \vert \alpha \right\rangle E_{\alpha}, \\ \\
t_{\bi{k}_{\parallel} \sigma}=-\frac{1}{f'^+_{\bi{k}_{\parallel} \sigma}}\sum_{\alpha} \left\langle \bi{k}_{\parallel} \sigma \vert \alpha \right\rangle E'_{\alpha},
\end{array}
\end{eqnarray}
where $f^{\pm}_{\alpha} = 1 \pm z_s Y_{\alpha}$ and $\delta_{ij}$ is the Kronecker's delta.  We
have also defined the quantities
$E_{\alpha}=C_{\alpha} f^-_{\alpha}+D_{\alpha} f^+_{\alpha}$ and
$E'_{\alpha}=-C_ {\alpha}f^+_{\alpha} \rme^{\rmi k_{\alpha z}
h}-D_{\alpha} f^-_{\alpha} \rme^{-\rmi k_{\alpha z} h}$, which (approximately)
represent the modal amplitudes of the electric field at the input
and output interfaces of the indentations, respectively.
The overlapping integral, $\left\langle  \bi{k}_{\parallel} \sigma \vert \alpha \right\rangle$, of the plane wave and the waveguide mode are analytically known for simple geometries of the indentations. For circular holes
these expressions can be found in \cite{RobertsJOSA87,Morse}.

Let us now impose the continuity of the tangential component of the magnetic field. As this boundary condition only holds at
the openings, we must project over waveguide modes, which form a complete set in that region. Plugging  $\rho_{\bi{k}_{\parallel} \sigma}$ and
$t_{\bi{k}_{\parallel}\sigma}$ \eref{eq:rhok} into the projection over waveguide modes, we
end up with the following set of equations for $E_\alpha$, $E'_\alpha$

\begin{eqnarray}
\label{eq:SE}
\begin{array}{l}
 \left( G_{\alpha \alpha}-\Sigma_{\alpha}\right)  E_{\alpha}+\sum_{\beta \neq \alpha} G_{\alpha \beta} E_{\beta} -G^V_{\alpha} E'_{\alpha} = I_{\alpha} \\ \\
\left( G'_{\gamma \gamma}-\Sigma_{\gamma}\right)  E'_{\gamma}+\sum_{\nu \neq \gamma} G'_{\gamma \nu} E'_{\nu} -G^V_{\gamma} E_{\gamma}= 0,
\end{array}
\end{eqnarray}

where
\begin{equation}
\label{Ialfa}
I_{\alpha}=2 \frac{Y_{\bi{k}^0_{\parallel} \sigma^0}}{f^+_{\bi{k}^0_{\parallel} \sigma^0}} \left\langle \alpha \vert  \bi{k}^0_{\parallel} \sigma^0 \right\rangle.
\end{equation}
\begin{equation}
\label{eq:Salpha}
 \Sigma_{\alpha}=\cases{ Y_{\alpha} \frac{ f^+_ {\alpha} \rme^{\rmi k_{\alpha z} h}+f^-_{\alpha} \rme^{-\rmi k_{\alpha z} h}}{ {f^+_{\alpha}}^2 \rme^{\rmi k_{\alpha z} h}-{f^-_{\alpha}}^2 \rme^{-\rmi k_{\alpha z} h}} & for a hole, \\
 Y_{\alpha} \frac{f^-_ {\alpha} \rme^{\rmi k_{\alpha z} h}+f^+_{\alpha} \rme^{-\rmi k_{\alpha z} h}}{f^+_{\alpha}  f^-_ {\alpha} (\rme^{\rmi k_{\alpha z}} - \rme^{-\rmi k_{\alpha z} h}) } & for a dimple. }
\end{equation}
\begin{equation}
\label{eq:GV}
 G^V_{\alpha}=\cases{  2 Y_{\alpha} \left[
{f^+_{\alpha}}^2 \rme^{\rmi k_{\alpha z}
h}-{f^-_{\alpha}}^2\rme^{-\rmi k_{\alpha z} h} \right]^{-1} & for a hole, \\
 0 & for a dimple. }
\end{equation}
\begin{equation}
\label{eq:green}
G_{\alpha \beta}=\sum_{\bi{k}_{\parallel} \sigma} \frac{Y_{\bi{k}_{\parallel} \sigma}}{f^+_{\bi{k}_{\parallel} \sigma}} \left\langle \alpha \vert \kappa \right\rangle \left\langle \kappa \vert \beta \right\rangle,
\end{equation}
and  $G'_{\gamma \nu}$  differs from $G_{\alpha \beta}$ only in the
constituent parameters, i.e.,  $G'_{\gamma \nu}$ is function of $\epsilon_3$ while $G_{\alpha \beta}$ depends on $\epsilon_1$.

The interpretation of these quantities is as follows: $I_{\alpha}$
takes into account the direct
illumination over object $\alpha$, $\Sigma_{\alpha}$ is
related to the bouncing back and forth of the waveguide fields
inside the indentations, $G^V_{\alpha}$ reflects the coupling
between EM fields at the two sides of a given indentation, while $G_{\alpha \beta}$ and
$G'_{\alpha \beta}$ are propagators that couple "objects" at the same side of the metal film.

If the system is periodic, with periodicity defined by $L_x$ and $L_y$,
$G_{\alpha \beta}$ is calculated
through the discrete sum over Bragg diffraction modes defined above.
For non-periodic systems, $L_x$ and $L_y$ define a fictitious super-cell and
the limit $L_x,L_y \rightarrow
\infty$ must be taken. This can be done analytically, simply replacing
 all sums in diffraction modes by integrals in $\bi{k}_{\parallel}$, i.e.
$\sum_{\bi{k}_{\parallel} \sigma} \rightarrow L_x L_y (2
\pi)^{-2}\sum_{\sigma} \int \rmd^2 \bi{k}_{\parallel}$. Notice that the
diverging factor $L_x L_y$ cancels out with the normalization factor of the plane waves.

Once the self-consistent ${E_{\alpha},E'_{\alpha}}$ are found, \eref{eq:rhok} gives the expansion coefficients
$\rho_{\bi{k}_{\parallel} \sigma}$ and  $t_{\bi{k}_{\parallel}
\sigma}$, which in turn can be used to obtain the EM fields everywhere (\ref{eq:EMFexpIII}
is devoted to derive the expressions for the
EM fields in region III). So far we have focused on in-plane components of the EM-fields. The z-components of the field can be readily computed using Maxwell equations.

For $z_s=0$ (perfect conductor case)
the expressions presented in this Appendix simplify to those derived in \cite{BravoPRL04}.
Notice however, that in \cite{BravoPRL04} the definitions of $I_\alpha, \, \Sigma_\alpha, \,
G_v, \, G_{\alpha \beta}$, and $G'_{\alpha \beta}$ contain an additional
multiplicative factor $\imath$. As explained in \cite{LMMJPC08}, this extra factor is useful  for drawing a parallelism with the "tight-binding" method employed in solid state physics, but it will be omitted here.

\section{EM fields in the region $z \geq h$}
\label{eq:EMFexpIII} Substituting the expression \eref{eq:rhok} of
$t_{\bi{k}_{\parallel} \sigma}$  in \eref{eq:EMFIII}, we can write 
the EM fields for $z \geq h$ as
\begin{eqnarray}
\label{eq:EMFr}
<r \vert \bi{E}' > = -\sum_{\alpha} \int \rmd^2 r'_{\parallel} 
\, \hat{\bi{G}}'_E ( r,r'_{\parallel}) <r'_{\parallel} \vert \alpha> E'_{\alpha}, \nonumber \\ \\
< r \vert -\bi{u}_z \times \bi{H}' > = -\sum_{\alpha} \int \rmd^2 r'_{\parallel}  \, \hat{\bi{G}}'_H ( r,r'_{\parallel} ) <r'_{\parallel} \vert \alpha >  E'_{\alpha}, \nonumber
\end{eqnarray}
where
\begin{eqnarray}
\label{eq:greenH}
\hat{\bi{G}}'_H ( r,r^{'}_{\parallel} ) =  \sum_{\bi{k}_{\parallel} \sigma} \frac{Y'_{\bi{k}_{\parallel} \sigma}}{f'^+_{\bi{k}_{\parallel} \sigma}} < r \vert \bi{k}_{\parallel} \sigma ><\bi{k}_{\parallel} \sigma \vert r'_{\parallel}> \rme^{i k'_z (z-h)},
\end{eqnarray}
and the corresponding function for the electric field,
$\hat{\bi{G}}'_E ( r,r^{'}_{\parallel})$, is obtained from
$\hat{\bi{G}}'_H (r,r^{'}_{\parallel})$ setting
$Y'_{\bi{k}_{\parallel} \sigma}=1$ in the numerator of the fraction given in the argument. Note (i) that the integration is
limited to the area of the indentations, and (ii) that we are not using the whole $3 \times 3$ matrix representation of the EM propagator but the $2 \times 2$ sub-matrix needed for projecting the EM fields in the bi-vectorial representation employed in this paper. An  expression similar to \eref{eq:EMFr} can be obtained for $z<0$.

The integral in the two-dimensional $\bi{k}_{\parallel}$-space can
be greatly simplified transforming into polar-coordinates and working on the proper  frame of reference. In order to do this, the $k_x$ axis should be oriented along the direction
$\bi{R}=\bi{r}_{\parallel}-\bi{r}'_{\parallel}$ rotating the
$\bi{k}_{\parallel}$-coordinate system by an angle $\theta$ around
the $z$-axis. The angular integration can be performed analytically by using the integral representation of the Bessel
function, $J_0(x)=(2 \pi)^{-1} \int^{2 \pi}_0 \exp(\rmi x \cos
\phi) \rmd \phi$. After some
straightforward algebra, the Green dyadic in the rotated coordinate frame is given by the
following diagonal matrix:
\begin{eqnarray}
\label{eq:greenH}
\fl \hat{\bi{G}}^{\theta}_H \left( R \right) =&
\frac{1}{4 \pi}\int k_{\parallel} \rmd k_{\parallel} \rme^{ \rmi k'_z (z-h)} \left\lbrace \mbox{diag} [ \Delta_{02},\Xi_{02} ]  \frac{Y'_{\bi{k}_{\parallel} p}}{f'^+_{\bi{k}_{\parallel} p}} + \mbox{diag} [\Xi_{02}, \Delta_{02}] \frac{Y'_{\bi{k}_{\parallel} s}}{f'^+_{\bi{k}_{\parallel} s}} \right\rbrace,
\end{eqnarray}
where $\Delta_{02}=J_0(k_{\parallel} R)-J_2(k_{\parallel} R)$ and
$\Xi_{02}=J_0(k_{\parallel} R)+J_2(k_{\parallel} R)$.

In general, $\hat{\bi{G}}^{\theta}_H$ must be computed numerically. Once this is done,
$\hat{\bi{G}}'_E ( r,r^{'}_{\parallel})$ is given by
$\hat{\bi{G}}'_E ( r,r^{'}_{\parallel})=\hat{\bi{\Theta}}^t(\theta) \,
\hat{\bi{G}}^{\theta}_H (R) \, \hat{\bi{\Theta}}(\theta)$,  where $\hat{\bi{\Theta}}(\theta)$ is the matrix representation of the operator for
a rotation by the angle $\theta$ around the $z$-axis:
\begin{eqnarray}
\hat{\bi{\Theta}}(\theta)=\left( \begin{array}{cc}
\cos \theta & \sin \theta \\
-\sin \theta & \cos \theta
\end{array}\right). \nonumber
\end{eqnarray}

\section{Transmitted power}
\label{sec:TW}
The power transmitted along the $z$-direction to
region III, $W'_z$, is obtained after integrating  the time-averaged
z-component of the Poynting vector, $\bi{S}_z$, at the output side
of the indentations,\footnote{The term $1/2$ is dropped for we
normalized by the incident flux containing the same factor.} i.e.
\begin{equation}
\label{eq:Wzprime}
W'_z
= \mbox{Re} [\left\langle -\bi{u}_z \times \bi{H} \vert \bi{E} \right\rangle ]={\sum_{\bi{k}_{\parallel} \sigma}}^p Y'_{\bi{k}_{\parallel} \sigma} \vert t_{\bi{k}_{\parallel} \sigma} \vert^2 = \sum_{\alpha \beta} E'^*_{\alpha} E'_{\beta} G^{W'}_{\alpha \beta},
\end{equation}
where
 \begin{equation}
\label{eq:greenjprime}
G^{W'}_{\alpha \beta}={\sum_{\bi{k}_{\parallel} \sigma}}^p \frac{Y'_{\bi{k}_{\parallel} \sigma}} {\vert f'^+_{\bi{k}_{\parallel} \sigma} \vert^2} \left\langle \alpha \vert \bi{k}_{\parallel} \sigma \right\rangle \left\langle
\bi{k}_{\parallel} \sigma\vert \beta \right\rangle.
\end{equation}
The superscript $p$ indicates that only propagating plane wave
solutions should be taken   into account.

Similarly, the energy power crossing the plane $z=0$ at the input side of the indentations in
region I, $W_z$, is
\begin{eqnarray}
\label{eq:Wz}
\fl W_z
=Y_{\bi{k}^0_{\parallel} \sigma^0}- {\sum_{\bi{k}_{\parallel} \sigma}}^p Y_{\bi{k}_{\parallel} \sigma} \vert \rho_{\bi{k}_{\parallel} \sigma} \vert^2  =\mbox{Re}\left[ \frac{f^{-*}_{\bi{k}^0_{\parallel} \sigma^0}}{f^+_{\bi{k}^0_{\parallel} \sigma^0}}\sum_{\alpha} E_{\alpha}I^*_{\alpha}  \right]- \sum_{\alpha \beta} E^*_{\beta} E_{\alpha} G^W_{\beta \alpha},
\end{eqnarray}
for the case of an incident plane wave, and
\begin{eqnarray}
\label{eq:WzSPP}
\fl W_z
=- {\sum_{\bi{k}_{\parallel} \sigma}}^p Y_{\bi{k}_{\parallel} \sigma} \vert \rho_{\bi{k}_{\parallel} \sigma} \vert^2  = - \sum_{\alpha \beta} E^*_{\beta} E_{\alpha} G^W_{\beta \alpha},
\end{eqnarray}
if the system is illuminated by a SPP.

The only difference between the definition for $G^W_{\alpha \beta}$ and $G^{W'}_{\alpha \beta}$
\eref{eq:greenjprime} is that the former depends on $\epsilon_1$ and the latter on $\epsilon_3$.

The energy power crossing a plane with constant $z$ in region II,
$W^{\scriptscriptstyle II}_z$, can be expressed as function of the expansion
coefficients $C_{\alpha}$ and $D_{\alpha}$:
\begin{eqnarray}
W^{\scriptscriptstyle II}_z =\sum_{\alpha} \left[ \mbox{Re}[Y_{\alpha}] \left( \vert C_\alpha \vert^2 \rme^{-2 {\small Im} [k_{\alpha z}] z}-\vert D_\alpha \vert^2 \rme^{ 2 {\small Im} [k_{\alpha z}] z} \right) \right.  \nonumber \\
\left.  + 2 \mbox{Im} [Y_{\alpha}] \left( C^*_{\alpha} D_{\alpha} \rme^{-2 \rmi {\small Re} [k_{\alpha z}] z} \right) \right].
\end{eqnarray}
If absorption is neglected, the previous expression simplifies to $W^{\scriptscriptstyle II}_z= \mbox{Re} \left[
\sum_{\alpha} E_{\alpha} E^{'*}_{\alpha} G^V_{\alpha} \right]$.

\section{Energy power scattered into SPPs}
\label{sec:SW}

The power scattered parallel to the metal plane is difficult to
compute due to multiple integrations over several components of the
EM fields \eref{eq:EMFr}. However, it can be shown that at a large
distance $R$ from the indentations, the main contribution to the propagator
given by \eref{eq:greenH}
comes from the divergent term $1/f^+_{\bi{k}_{\parallel} p}$. The propagator presents
a pole at the SPP wavevector which, within the SIBCs, occurs when \footnote{In this part
of the text we do not make a distinction between the region I with
dielectric constat $\epsilon_1$ and the region III with
$\epsilon_3$. A generic value of $\epsilon$ is used instead since the
deduced formulas are valid for both regions.}
\begin{equation}
\label{eq:kspp}
k^{\mbox{{\tiny spp}}}_{\parallel
}=k_\omega \sqrt{\epsilon(1-\epsilon/\epsilon_M)}.
\end{equation}
We also assume that
the rest of the integrand, $F(k_{\parallel})$, is practically
constant in the region of the upper complex semi-plane close to this
SPP pole, and that we can replace the exact result of the integral
by its residue, i.e. $\int^{\infty}_0 \rmd k_{\parallel}
F(k_{\parallel}) [k_z+z_s \sqrt{\epsilon} k_\omega]^{-1} \approx 2 \pi
\rmi \epsilon z_s k_\omega F(k^{\mbox{{\tiny
spp}}}_{\parallel})/k^{\mbox{{\tiny spp}}}_{\parallel}$.

Special care must be taken when integrating the Bessel function, as this function diverges
for large arguments both in the lower and upper complex semi-planes. This problem can be
solved by using the identity $2J_n(x)=H^{(1)}_n+H^{(2)}_n$, where $H^{(1)}_n$ and
$H^{(2)}_n$ are the n-th order Hankel functions of first and second kind, respectively.
The part of the argument containing
$H^{(1)}_n$ must be integrated in the upper complex semi-plane,
while $H^{(2)}_n$ has to be integrated in the lower semi-plane. Only the integration
over the upper complex semi-plane encloses the SPP pole. The integral in the lower semi-plane vanishes identically. As a result, and using the
asymptotic expression $H^{(1)}_n(x>>1) \approx \sqrt{2/(\pi x)}
\exp[\rmi (x-n \pi/2-\pi/4)] $, we arrive at the long-distance
asymptotic expression for the propagator governing the launching of SPP
\begin{eqnarray}
\begin{array}{l}
 G^{\mbox{{\tiny spp}}}_H(R)=\gamma \, \mbox{diag}[1,0][k^{\mbox{{\tiny spp}}}_{\parallel} R]^{-1/2} \exp[\rmi (k^{\mbox{{\tiny spp}}}_{\parallel} R+k^{\mbox{{\tiny spp}}}_z z)], \\
G^{\mbox{{\tiny spp}}}_E(R)=Y^{-1}_{\mbox{{\tiny spp}}} G^{\mbox{{\tiny spp}}}_H(R),
\end{array}
\end{eqnarray}
where $\gamma=\rmi \epsilon^2 k^2_\omega z_s \rme^{-\rmi\pi/4} [2 \pi]^{-1/2}$  and $Y_{\mbox{{\tiny spp}}}=-z^{-1}_s$ is the admittance evaluated at $k^{\mbox{{\tiny spp}}}_{\parallel}$.
Comparison with numerical results shows that, the previous expression is an excellent
approximation in the optical regime for good metals (like Au and Ag) already for $R \gtrsim 2 \lambda$, despite being a long-distance
asymptotic result.

Up to this point we have assumed a single indentation is placed at the
origin of the system of reference . We now consider  the case of an arbitrary number of
indentations,
located at arbitrary positions $\bi{R}_{\alpha} = |\bi{R}_{\alpha}| \, \bi{u}_{\alpha}$.
In order to compute the long-distance asymptotic expression, at a given observation point $R$, of the SPP field launched by the
indentations,
we only take into account terms of order $\Or (R_{\alpha}/R)$. Then
$\vert \bi{R}-\bi{R}_{\alpha}-\bi{r}' \vert \approx R-\bi{R}_{\alpha}
\cdot \bi{u}_R-\bi{r}' \cdot \bi{u}_R$, where $\bi{u}_R$ is a unitary vector directed from the indentation to the point $R$ and $\bi{r}'$ is the local coordinate inside the hole. 
The SPP EM fields can be computed after integrating the
propagator over the different areas occupied by the indentations, as indicated in
\eref{eq:EMFr}.

After performing the integration, we obtain for the non-vanishing components of the
SPP field:
\begin{eqnarray}
\label{eq:Fspp}
\begin{array}{l}
 H^{\mbox{{\tiny spp}}}_{\theta}(\bi{R})=\gamma   [k^{\mbox{{\tiny spp}}}_{\parallel} R]^{-1/2} \exp[\rmi (k^{\mbox{{\tiny spp}}}_{\parallel} R+k^{\mbox{{\tiny spp}}}_z z)] \,
f(\bi{u}_R) , \\
E^{\mbox{{\tiny spp}}}_R(\bi{R})=H^{\mbox{{\tiny spp}}}_{\theta}(\bi{R})/Y_{\mbox{{\tiny spp}}}, \\
E^{\mbox{{\tiny spp}}}_z(\bi{R})=V H^{\mbox{{\tiny spp}}}_{\theta}(\bi{R})/Y_{\mbox{{\tiny spp}}},
\end{array}
\end{eqnarray}
where $V=-k^{\mbox{{\tiny spp}}}_{\parallel}/k^{\mbox{{\tiny spp}}}_z$.

The function $f(\bi{u}_R)$ gives the angular dependence of the total launched SPP field, and accounts for the interference of
the SPPs launched by all the different waveguide modes in all the different indentations
\begin{equation}
f(\bi{u}_R)=\sum_{\alpha} f_{\alpha}( \mathbf{u}_R )
\rme^{-\rmi k^{\mbox{{\tiny spp}}}_{\parallel} \left(
\bi{R}_{\alpha} \cdot \bi{u}_R\right) } E_{\alpha},
\end{equation}
where the angular dependence of the SPP field scattered by a given normalized waveguide mode (with in-plane electric field given by $\bi{E}_{\alpha}(\bi{r}')$)
is provided by
\begin{equation}
f_{\alpha}( \bi{u}_R ) = \int \rmd \bi{r}' \rme^{-\rmi
\bi{k}^{\mbox{{\tiny spp}}}_{\parallel} \cdot \bi{r}'}
\bi{E}_{\alpha} (\bi{r}')\cdot \bi{u}_R.
\end{equation}

The SPP scattered power,
$W_{\scriptscriptstyle spp}$, is computed by integrating
the time-averaged radial
component of the Poynting vector, $\mathbf{S}_{\scriptscriptstyle
spp} = \mbox{Re} [\bi{E}^{\scriptscriptstyle \rm spp} \times
\bi{H}^{\scriptscriptstyle \rm spp *}]$ on the lateral side of an
imaginary cylinder of radius $R$, i.e $W_{\scriptscriptstyle spp}=\int^{\infty}_0 \rmd z \int^{2 \pi}_0 R \rmd \theta
\bi{S}_{\scriptscriptstyle spp} \cdot \bi{u}_R$, ending up with

\begin{eqnarray}
\label{eq:Wspp}
W_{\scriptscriptstyle spp} =  \Omega_{\scriptscriptstyle spp} \exp (-2 \vert \mbox{Im} [k^{\scriptscriptstyle \rm spp}_{\parallel} ] \vert R),
\end{eqnarray}
where
\begin{equation}
\label{eq:Ospp}
 \Omega_{\scriptscriptstyle spp}=\epsilon^2 \frac{\vert z_s \vert k^2_0  \bar{f}_2}{2}  \left[ 1+ \frac{ \mbox{Re}[k^{\scriptscriptstyle \rm spp}_z]^2}{\mbox{Im} [k^{\scriptscriptstyle \rm spp}_z]^2} \right]^{1/2} \left[ 1+ \frac{\mbox{Im}[k^{\scriptscriptstyle \rm spp}_{ \parallel}]^2}{\mbox{Re} [k^{\scriptscriptstyle \rm spp}_{\parallel }]^2} \right]^{-1/2} ,
\end{equation}
and $\bar{f}_2=(2 \pi)^{-1} \int^{2 \pi}_0 \vert f \left( \mathbf{u}_R \right) \vert^2 d \theta$.

This is the result integrated over all angles. Clearly, the quantity $\vert f(\bi{u}_R) \vert^2$ determines the angular distribution of the SPP scattered power.

For a lossless material $z_s=i \mbox{Im} [z_s]$, $\mbox{Im} [k^{\scriptscriptstyle
spp}_{\parallel}]=0$, and $\mbox{Re} [k^{\scriptscriptstyle spp}_z]=0$. Therefore  $W_{\scriptscriptstyle spp}$ is independent of  $R$;
otherwise $W_{\scriptscriptstyle spp}$ decays as function of $R$.
For lossy systems $\Omega_{\scriptscriptstyle spp}$ still provides the total energy scattered into the
plasmon channel; the dependence with distance to the scatterers of $W_{\scriptscriptstyle spp}$ follows the decay of the energy stored in the SPP, due to absorption by the metal.

Finally, we adopt the following normalization for the scattering problem related to the
out-coupling  of a SPP into propagating radiation. We assume a SPP propagating at the
metal-dielectric interface is incident on the system composed of
either  hole or dimples, at an angle $\theta_0$ taken from the
$x$-axis, see \fref{fig:scheme}(b). The incident EM fields are given
by
\begin{eqnarray}
\label{eq:Finc}
\begin{array}{l}
 \bi{E}_{\scriptscriptstyle inc}=\rme^{\rmi \bi{k}^{\scriptscriptstyle spp}  \cdot \bi{r}} (\cos \theta_0,\sin \theta_0,V)^t/Y^{\scriptscriptstyle spp}, \\
 \bi{H}_{\scriptscriptstyle inc}=\rme^{\rmi \bi{k}^{\scriptscriptstyle spp}  \cdot \bi{r}} (-\sin \theta_0,\cos \theta_0,0)^t,
\end{array}
\end{eqnarray}
from which we derive the time-averaged Poynting vector along
an arbitrary radial direction $\bi{r}$, i.e. $S^{\scriptscriptstyle inc}_{\bi{r}}=\rme^{-2 \vert {\small Im} [\bi{k}^{\scriptscriptstyle spp}
\cdot \bi{r}] \vert} \mbox{Re} [\bi{k}^{\scriptscriptstyle spp}  \cdot
\bi{u}_{\bi{r}}]/(\epsilon k_\omega)$. We define the incident power,
$W^{\scriptscriptstyle spp}_{\scriptscriptstyle \rm inc}$,
integrating $S^{\scriptscriptstyle inc}_{\bi{r}}$ in the rectangular area at $x=0$, which extends along the
side of the defect in the $y$-direction, and along $z>0$ in the
$z$-direction. For a SPP impinging onto a single circular indentation (with radius $r$) we obtain
\begin{equation}
W^{\scriptscriptstyle spp}_{\scriptscriptstyle \rm inc}=\int_{2r}\rmd y \int^{\infty}_0 \rmd z S^{\scriptscriptstyle inc}_{\bi{r}} \cdot \bi{u}_x=\frac{r}{\epsilon k_\omega} \frac{\mbox{Re} [k^{\scriptscriptstyle \rm spp}_{\parallel}]}{\vert \mbox{Im} [k^{\scriptscriptstyle \rm spp}_z] \vert}.
\end{equation}

\section*{References}

\end{document}